# Phase-slip flux qubits

**J E Mooij and C J P M Harmans**
Kavli Institute of Nanoscience, Delft University of Technology,
2628 CJ Delft, The Netherlands

E-mail: j.e.mooij@tnw.tudelft.nl

**Abstract.** In thin superconducting wires, phase-slip by thermal activation near the critical temperature is a well-known effect. It has recently become clear that phase-slip by quantum tunnelling through the energy barrier can also have a significant rate at low temperatures. In this paper it is suggested that quantum phase-slip can be used to realize a superconducting quantum bit without Josephson junctions. A loop containing a nanofabricated very thin wire is biased with an externally applied magnetic flux of half a flux quantum, resulting in two states with opposite circulating current and equal energy. Quantum phase-slip should provide coherent coupling between these two macroscopic states. Numbers are given for a wire of amorphous niobium-silicon that can be fabricated with advanced electron beam lithography.

## 1. Introduction

The phase difference of the order parameter over a thin, narrow superconducting wire can change in units of $2\pi$ by means of a discrete process that is localized in space and time. The amplitude of the order parameter becomes smaller over a few coherence lengths, touches zero in one spot at one instant and then recovers its initial value. After this phase-slip event, the phase difference has changed by $2\pi$. There is an energy barrier against phase-slip, because the temporarily reduced order parameter in the phase-slip volume leads to a loss of condensation energy. The phase-slip process has been studied in detail at temperatures near the critical temperature, where thermal activation allows crossing over the energy barrier [1]. Phase slip leads to dissipation for currents below the critical pair-breaking value. For one-dimensional superconductors it is the equivalent of vortex crossing in two- and three-dimensional systems. Phase-slip is also possible at zero temperature by means of quantum tunnelling. To observe this quantum phase-slip, one needs extremely weak superconducting wires. Over the years, several experiments were performed where significant deviations from thermal activation were observed at low temperatures [2,3]. It was hard to compare these data with theoretical predictions, because quantitative calculations were not available. Recently, convincing experimental evidence for quantum phase-slip at zero temperature was presented by the Harvard group of Tinkham and collaborators [4,5], who fabricated extremely thin wires by covering a suspended carbon nanotube with a film of amorphous Mo-Ge. New microscopic theoretical calculations were performed by Zaikin and collaborators [6,7]. Experimental results on a series of Mo-Ge wires were compared with this theory and reasonable agreement was found.

So far, phase-slip has been studied in wires that were connected to an external current source. Each event is connected with an energy transfer $I\Phi_0$, where $I$ is the current and $\Phi_0 = h/2e$ is the flux quantum. Induced by thermal activation or by quantum tunnelling, phase slip is manifested as a stochastic process that leads to dissipation. However, if the concept of quantum phase-slip is valid, it should also be possible to use the effect as a coherent coupling between two distinct macroscopic quantum states in a superconducting quantum circuit. We propose to study the tunnel strength for phase-slip transitions in a closed superconducting loop that contains a weak wire, while a magnetic flux of half a flux quantum is applied to the loop. The energy after macroscopic quantum tunnelling is the same but the persistent current is reversed. Quantum phase-slip should lead to a quantum superposition of the two opposed current states. One would obtain superconducting quantum bits, more in particular flux qubits, in which Josephson tunnel junctions are replaced by a weak superconducting wire. Using the established techniques for experiments on junction-based flux qubits, quantum phase-slip can be studied under optimally coherent conditions. The fundamental question concerning superconducting long-range order in an increasingly weak one-dimensional wire can be addressed without the use of dissipative processes. The application as a phase-slip qubit may have clear practical advantages over junction-based flux qubits as will be discussed at the end of this paper.

## 2. Wires and loops

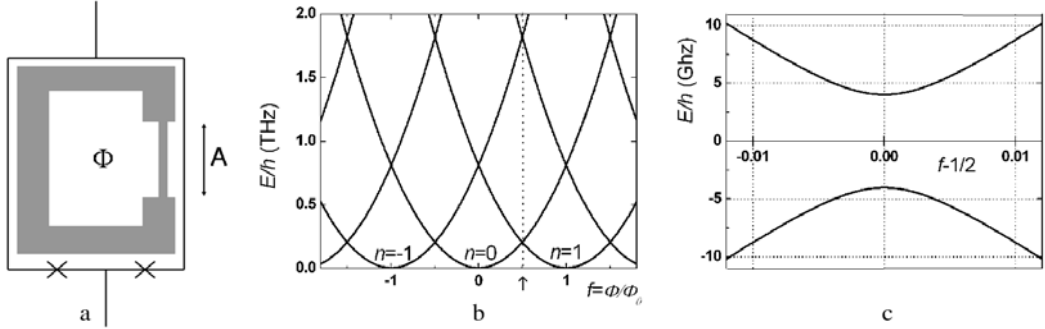

Figure 1. a. Schematic lay-out of the weak wire in a closed superconducting loop. A SQUID with two Josephson junctions (crosses) is used to measure the qubit state. b. Energy levels as a function of applied flux for different fluxoid numbers, according to equation (6) for a loop with $L_k$ = 4 nH. A phase-slip event changes the fluxoid number $n$. The arrow indicates the operating point at $f$ = ½. c. Lowest two energy levels of phase-slip qubit with $I_p$=0.25 µA and $\Gamma_{qps}/2\pi$= 4 GHz near $f$ = ½.

Consider a long superconducting wire with length $L$ and normal state resistance $R_n$ at temperature zero. The transverse dimensions are much smaller than the coherence length $\xi$= $(\xi_o \ell)^{\frac{1}{2}}$, where $\xi_o$ is the BCS coherence length and $\ell$ the electronic mean free path ($\ell << \xi_o$). The kinetic inductance $L_k$ quantifies the superconducting current response to a gauge-invariant phase difference $\gamma$, according to

$$I_s = \frac{\Phi_o}{2\pi} \frac{1}{L_k} \gamma \qquad (1)$$

The increase of energy due to current and phase difference is

$$E_\gamma = \frac{1}{2}\left(\frac{\Phi_o}{2\pi}\right)^2 \frac{1}{L_k} \gamma^2 \qquad (2)$$

Within a Ginzburg-Landau-like description, with the density of superconducting electrons derived from BCS/Gorkov theory for dirty superconductors at $T$=0, one finds

$$L_k = a \frac{\hbar R_n}{k_B T_c} \qquad (3)$$

where $T_c$ is the critical temperature and $a$ is a numerical constant (an estimate is $a \approx 0.14$). The maximum (pair-breaking) supercurrent in the wire can be estimated in a similar approach:

$$I_o = b \frac{k_B T_c}{e R_\xi} \qquad (4)$$

Here $R_\xi$ is the resistance of a length of wire equal to $\xi$, so that $R_\xi/R_n = \xi/L$. The prefactor $b$ is a constant of order unity, an estimated value is $b \approx 1.3$.

Now bend the wire to form a closed superconducting loop. A wider section can be included with negligibly low kinetic induction (figure 1a). An external source applies a magnetic flux $\Phi_{ext} = f\Phi_0$ through the loop. The induced current in the loop $I$ generates a flux $I L_g$ where $L_g$ is the geometric inductance of the loop. Single-valuedness of the order parameter requires that

$$\gamma = 2\pi f - 2\pi \frac{L_g I_s}{\Phi_o} - 2\pi n = 2\pi f - \frac{L_g}{L_k}\gamma - 2\pi n$$

with $n$ an integer. The value of $n$ can, after cool-down, only be changed by a phase-slip event. In practice, the kinetic inductance of the wires that exhibit quantum phase-slip is of order nH, while the geometric inductance is of order 10 pH. In calculating $\gamma$, the flux generated by the current can be ignored. As a consequence we can write

$$\gamma = 2\pi(f - n) \tag{5}$$

$$E = \frac{1}{2}\frac{\Phi_o^2}{L_k}(f - n)^2 \tag{6}$$

Figure 1b gives an example of the levels for various values of $n$ when $L_k = 4$ nH. In the immediate neighbourhood of $f = ½$ where the lines for $n=0$ and $n=1$ cross, defining $\delta f = f - ½$, one finds

$$E_{1,2} = \frac{1}{2}\frac{\Phi_o^2}{L_k}\left(\frac{1}{4} \pm \delta f\right) \tag{7}$$

The next higher energy levels for $f \approx ½$ are at a large distance, about $\Phi_o^2/L_k$ higher.
The 'classical' circulating persistent currents are $I_{1,2} = \pm I_p$ with $I_p = \Phi_o/2L_k$. For $L_k$ equal to 4 nH, $I_p$ is 0.25 µA. This is similar to values of persistent currents in Josephson junction flux qubits.

If the quantum phase slip rate is $\Gamma_{qps}$, the wire flux qubit has the following Hamiltonian (viewed as a pseudospin in the basis where the persistent currents create magnetization in the z-direction):

$$H = \delta f \Phi_o I_p \sigma_z + \hbar \Gamma_{qps} \sigma_x \tag{8}$$

$\sigma_z$ and $\sigma_x$ are Pauli spin operators. The qubit energy splitting is

$$dE = \pm\sqrt{(\delta f \Phi_o I_p)^2 + (\hbar \Gamma_{qps})^2} \tag{9}$$

In figure 1c the qubit levels near $f = ½$ are plotted for $I_p = 0.25$ µA and a phase-slip rate $\Gamma_{qps}/2\pi = 4$ GHz.

**3. Phase-slip rates**
To estimate the rate of quantum phase slip in the wire at low temperature we follow the discussion by Lau et al. [5], based on a heuristic extension to low temperatures of the well-established theory for thermally activated phase-slip near the critical temperature [1]. The barrier height for a phase-slip event at zero temperature in their description is

$$E_B = 0.83\frac{R_q}{R_\xi}k_B T_c \tag{10}$$

where $R_q = h/4e^2 = 6.5$ kΩ. Counterintuitively, in this expression a longer coherence length leads to a lower barrier, as $R_\xi = R_n'\xi$. This is due to the fact that the condensation energy density scales with $\xi_0^{-2}$ and all other factors combine to give the elegant form above. For a given strip $R_n'$ and $T_c$ are easily determined, but the coherence length is not well known in the relevant materials.

The attempt frequency is

$$\Omega = 0.57 \frac{L}{\xi} \sqrt{0.83 \frac{R_q}{R_\xi}} \frac{1}{\tau_{GL}} \tag{11}$$

with $\tau_{GL}=\pi\hbar/8k_BT_c$ the Ginzburg-Landau time at zero temperature ($1/\tau_{GL}=1.4\Delta(0)/\hbar$ in terms of the BCS gap). The phase-slip rate amounts to

$$\Gamma_{qps} = 1.5c \frac{L}{\xi} \sqrt{\frac{R_q}{R_\xi}} \frac{k_B T_c}{\hbar} \exp\left(-0.3d \frac{R_q}{R_\xi}\right) \tag{12}$$

$c$ and $d$ are unknown constants of order unity that express the uncertainties in the derivation. Lau et al. could fit the resistance due to driven phase-slip in a large set of experimental wires, with known values of their resistance per unit length, quite accurately to equation (12). From their fit, one would induce that $d=1$ within the uncertainty of knowing the precise value of $\xi$.

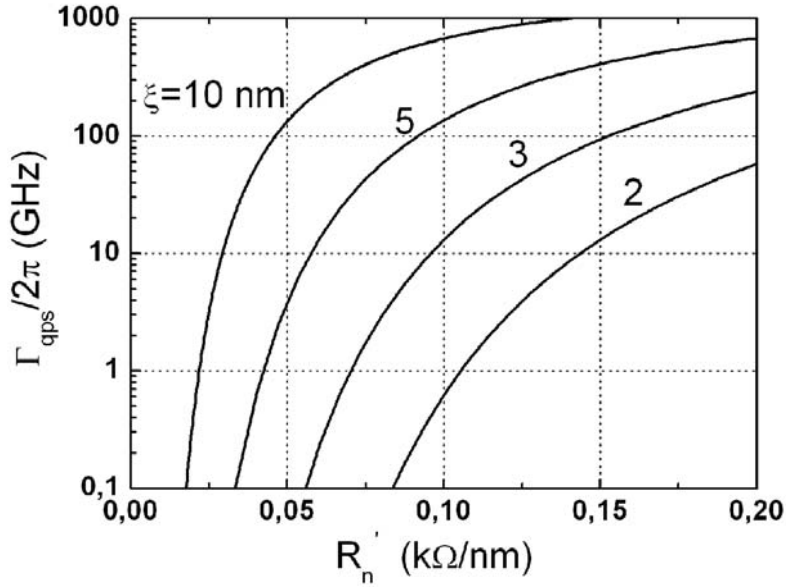

Figure 2. Quantum phase-slip rate as a function of resistance per unit length for four values of $\xi$, according to equation (12). Assumptions: $L=50$ nm, $T_c=1.2$ K, $c=d=1$.

In figure 2 we plot $\Gamma_{qps}/2\pi$ versus $R_n'$ as calculated from (12) for various values of $\xi$. Desired values of $\Gamma_{qps}$ for qubits are in the range of 1-10 GHz. The figure makes clear that $R_\xi=R_n'\xi$ has to be around 0.3 k$\Omega$, and also illustrates how sensitively the quantum tunnelling depends on the precise value of $R_\xi$. For practical purposes as a qubit one must control $R_\xi$, and therefore the line width, to within a few percent.

**4. Practical wire qubits**
Clearly superconducting wires with very high resistance are needed. Can such wires in circuits be fabricated with established clean room techniques? In the following we analyze realistic values for a wire of amorphous Nb-Si. These wires can be fabricated from very thin amorphous films that are homogeneous to below a nanometer scale. The material is similar to the Mo-Ge of [8] and [4]. In our group, we studied a Nb-Si wire of 130 nm width that clearly exhibited deviations from thermal activation in its low temperature resistance [3]. Electron beam lithography at this time allows fabrication of wires with width down to 10 nm, starting from ultrathin films. Co-sputtered films of Nb-Si with composition around Nb50%-Si50% showed a resistivity of about 2.0 $\mu\Omega$m. The critical temperature changed with thickness and with composition [9]. In the following we choose as an example a Nb-Si film with composition 42% Nb-58%Si and a nominal thickness of 2.5 nm, to be the starting material. The film showed a sharp superconducting transition at 1.2 K, and metallic temperature dependence above $T_c$. The sheet resistance was 1 kΩ/square. A 10 nm wide wire of this material will have a resistance $R_n' = 0.1$ kΩ/nm. The electronic mean free path is estimated to be $\ell = 0.4$ nm. The BCS coherence length is not known, but given the low critical temperature it is likely to exceed the value for Nb which is 20 nm. This would lead to values of $\xi$ that could range from 3 to 8 nm. For that range of $\xi$ and for $R_n' = 0.1$ kΩ, equation (12) gives rates $\Gamma_{qps}/2\pi$ ranging from 4 to 60 GHz. One cannot draw firm conclusions, but it seems there is a good chance that with practical wires one can achieve a quantum phase-slip rate in the GHz range.

We now discuss the practical design of a phase-slip flux qubit that is based on a wire of this Nb-Si material. The phase-slip rate depends exponentially on the resistance in a length $\xi$ of the wire; the length of the wire has only linear influence. Given the desired phase-slip rate and the corresponding desired resistance per unit length near 0.1 kΩ/nm, one can use the length of the wire to optimize the energy and the persistent current level. The example discussed above with $L = 50$ nm leads to $L_k = 4$ nH and $I_p = 0.25$ μA, a convenient level for SQUID measurement.

The phase-slip qubit potentially has two strong advantages. The first is connected with 1/f-type noise. In the existing superconducting qubits, based on Josephson junctions, charge noise and critical current noise are very significant sources of decoherence, usually limiting performance [12,13]. The phase-slip qubit is highly insensitive to charge noise, as it has no islands between tunnel junctions. The equivalent of critical current noise would be fluctuations of the energy barrier or the loop inductance. Phase-slip is an exponential process as is electron tunnelling through oxide barriers. However, the barrier is determined by the geometry and superconducting properties on the scale of the coherence length. Ionization or microscopic motion of a single atom will not influence the tunnel barrier by a significant amount. The loop inductance is even more macroscopic. The qubit will be sensitive to flux noise in the same way as junction flux qubits. Trapping and detrapping of vortices in wider parts of the qubit loop should be kept in mind, but is unlikely with the smooth homogeneous films of Nb-Si.

The other advantage of the phase-slip qubit is the large distance of more than 500 GHz between the qubit energy levels and the next higher states of the loop. In superconducting phase and charge/phase qubits, the next higher levels are close on the scale of the qubit splitting. Strong excitation may lead to leakage of quantum information. In junction flux qubits, the distance to the next higher level is a few qubit splittings away. In the phase-slip qubit, extremely fast excitation can be allowed.

**5. Conclusions**
Quantum phase-slip in superconducting wires that can be fabricated with advanced electron beam lithographic techniques can have a rate of the order of several GHz. Superconducting flux quantum bits can be developed where Josephson junctions have been replaced by a superconducting quantum wire. The two states of the wire qubit will be very well separated from the higher levels. The wire qubit will likely have much lower 1/f noise than the presently used superconducting qubits.

Acknowledgements. We thank M. Tinkham, G. Schön and A. Zaikin for very helpful discussions. This research is supported by FOM and by EU through SQUBIT2.